\documentclass[aps,prd,floats,graphicx,twocolumn]{revtex4}
\usepackage{graphicx}
\newcommand{\beq}{\begin{equation}}
\newcommand{\eeq}{\end{equation}}

\begin{document}


\title{Cosmic distance-duality as probe of exotic physics and acceleration}

\author{Bruce A. Bassett$^\natural$ and Martin Kunz$^\S$}

\address{$^\natural$ Department of Physics, Kyoto University, Kyoto, Japan  \&  Institute of Cosmology and Gravitation, 
University of Portsmouth, Portsmouth~PO1~2EG, UK
\\
$^\S$ Astronomy Centre, University of Sussex, Brighton, UK}
\date{\today}
\begin{abstract}
In cosmology, distances based on standard candles (e.g. supernovae) and standard rulers (e.g. baryon oscillations) agree as long as three conditions are met: (1) photon number is conserved, (2) gravity is described by a metric theory with (3) photons travelling on unique null geodesics. This is the content of distance-duality (the reciprocity relation) which can be violated by exotic physics. Here we  
analyse the implications of the latest cosmological data sets for distance-duality. While broadly in agreement and confirming acceleration we find a 2-sigma violation caused by excess brightening of SNIa at $z > 0.5$, perhaps due to lensing magnification bias. This brightening has been interpreted as evidence for a late-time transition in the dark energy but because it is not seen in the $d_A$ data we argue against such an interpretation. Our results do, however, rule out significant SNIa evolution and extinction: the ``replenishing" grey-dust model with no cosmic acceleration is excluded at more than 4-sigma despite this being the best-fit to SNIa data alone, thereby illustrating the power of distance-duality even with current data sets. 
\end{abstract}
\maketitle

\section{Introduction}
In 1933 Etherington \cite{eth,ellis,SEF} proved a beautiful and general duality that implies that distances in cosmology based on a metric theory of gravity are unique: whether one uses the apparent luminosity of standard candles (yielding the luminosity distance, $d_L(z)$) or the apparent size of standard rulers (the angular-diameter distance $d_A(z)$),  does not matter since they are linked by distance-duality\footnote{We use this term for clarity when referring specifically to the relation between $d_A$ and $d_L$ instead of the term `reciprocity' used in the general relativity literature to refer to the purely geometric relation between up-going and down-going null geodesic bundles and which makes no reference to $d_L$ \cite{SEF}.}:
\beq
\frac{d_L(z)}{d_A(z)(1+z)^2} = 1\,.
\label{recip}
\eeq
where $z$ is redshift. 
Distance-duality holds for general metric theories of gravity in any background (not just FLRW) in which photons travel on unique null geodesics and is essentially equivalent to Liouville's theorem in kinetic theory. While it is impervious to gravitational lensing (for infinitesimal geodesic bundles) it depends crucially on photon conservation. 
Our aim in this paper is to discuss how distance-duality may become a powerful test of a wide range of both exotic and fairly mundane physics and to present a general analysis of what constraints on violations of distance-duality arise from current data as well as critically analysing the conclusions drawn from recent type-Ia supernovae data \cite{riess} (also discussed in the appendix).

To test distance-duality we use the latest type Ia supernovae (SNIa) data \cite{riess,knop,barris,tonry} as a measure of the luminosity distance, $d_L(z)$ \cite{PS}. This data includes a significant number of $z>1$ observations. Our estimates of the angular-diameter distance, $d_A(z)$, come from FRIIb radio galaxies \cite{daly1,daly2}, compact radio sources \cite{crg,JD,jackson} and X-ray clusters \cite{allen}. It is important to remember that some of this data predated the discovery of acceleration by SNIa and that there are now completely independent, indirect, estimates of $d_A$, e.g. from analysis of the 2QZ quasar survey \cite{outram} (giving $\Omega_{\Lambda}=0.71^{+0.09}_{-0.17}$) and strong lensing from a combination of the CLASS and SDSS surveys with a maximum likelihood value of $\Omega_\Lambda = 0.74-0.78$ \cite{keeton}, in good agreement with estimates from radio sources.

All these data sets broadly agree with an accelerating, high-$\Omega_{\Lambda}$ cosmology. Nevertheless, there are a few observations in disagreement with the accelerating `concordance' model (e.g. \cite{blanchard}), there are suggestions that SNIa may suffer from significant extinction \cite{mrr}, evolution \cite{drell} or axion-photon mixing \cite{csaki}. There are also radical alternatives to general relativity, such as  MOND \cite{mond}. Distance-duality gives us a way to test all of these possibilities. 

\section{Distance-duality violations}
Since our aim in this paper is to promote distance-duality as a powerful test of fundamental physics it seems appropriate to begin by describing some phenomena that could be detected through violations of distance-duality. 
The most radical violations would arise from deviations from a metric theory of gravity or in cases where photons do not travel on (unique) null geodesics (e.g. torsion or birefringence). Other interesting possibilities include variation of fundamental constants such as $G$, but we do not discuss any of these possibilities here because they are either already tightly constrained, difficult to give predictions for or are too implausible given current prior beliefs about gravity. 

Instead we restrict ourselves to phenomena that may reasonably occur given our understanding of particle physics or astrophysics. We will also not discuss obvious possible sources of violation such as unaccounted for systematic error or biases in estimates of either $d_L$ and $d_A$. While this would be obvious first place to look to explain a violation of distance-duality we consider it to be trivial and hence will not discuss it further except when we put limits on the size of such effects later using distance-duality.

\subsection{Photon number violation}

Perhaps one of the most likely sources of duality-violation is non-conservation of photon number. This could have a mundane origin (scattering from dust or free electrons) or an exotic origin (e.g. photon decay or photon mixing with other light states such as the dark energy, dilaton or axion \cite{csaki,BK}). However, all of these effects tend to reduce the number of photons in a light bundle and therefore {\em reduce} the apparent luminosity of a source. If unaccounted for, this dimming makes the source appear more distant, i.e. increases $d_L$. Since $d_A$ is typically unaffected (or negligibly altered) by such effects, this rather generally implies that the ratio in equation (1) becomes greater than unity. The case of axion-photon mixing has been studied in \cite{BK} and the results there show that this type of dimming cannot obviate the need for cosmic acceleration. 

\begin{figure}[tbp]
\includegraphics[width=85mm]{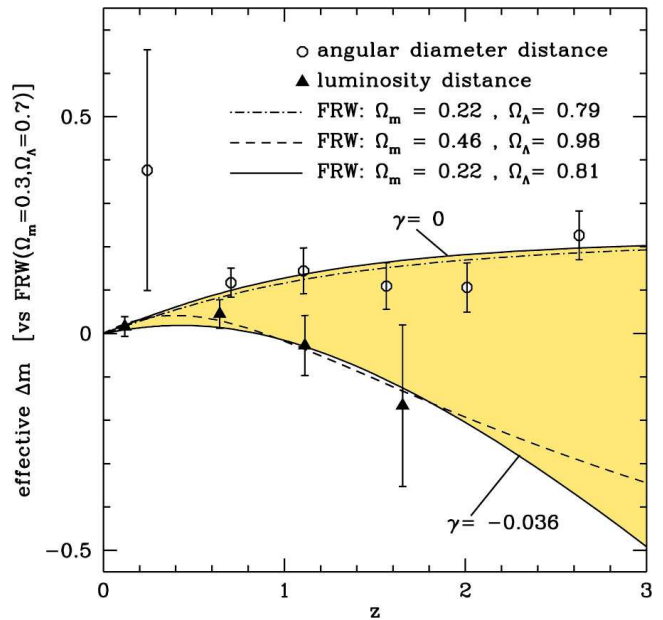} \\
\caption{{\bf Graphic evidence for violation of distance-duality.} The binned data for $d_L(z)$ (triangles, SNIa) and $d_A(z)$ (circles) are shown in equivalent magnitudes relative to the flat concordance model ($\Omega_{\Lambda}=0.7,\Omega_m=0.3$) with $1\sigma$ error bars. They should coincide if distance-duality holds but they differ significantly at $z>0.7$. The dashed curves are the best-fit FLRW models to the $d_A(z)$ data (top) and $d_L(z)$ (bottom) separately with no loss of photons ($\gamma = 0)$. The solid curves have the same underlying FLRW model ($\Omega_\Lambda=0.81$, $\Omega_m=0.22$) but the lower curve includes the best-fit brightening  ($\gamma = -0.036$, see eq. 2) with $\alpha=-2, \beta=1$. Since the violation of distance-duality increases exponentially when $\gamma \neq 0$ more high redshift data and/or smaller error bars will significantly improve constraints.}
\label{fig1}
\end{figure}

We can parametrise scattering or loss of photons by studying the photon Boltzmann equation integrated over frequency allowing for a collision functional: 
\beq
\dot{n}_{\gamma} + 3H n_{\gamma} = -2\gamma H_0 (1+z)^{1-\alpha} n_{\gamma}
\eeq
$n_{\gamma}$ is the number density of photons and $\gamma, \alpha$ are constants that control the scattering/decay cross-section of the photon. $H_0$ is the current value of the Hubble constant. $\alpha = -2$ corresponds to a scattering cross-section $\propto \rho_{cdm} \propto (1+z)^3$, as in the case of Compton scattering from free-electrons. The case of photon decay corresponds to $\alpha=1$. $\gamma > 0$ implies loss of photons.  In fact $\gamma \neq 0$ leads to a violation of distance-duality that grows roughly exponentially with redshift, see Fig (\ref{fig1}).

For the sake of generality we also consider power-law deformations of distance-duality that parametrise our ignorance about the effects of more exotic physics. This yields a 3-parameter $(\alpha,\beta,\gamma)$ extension of equation (\ref{recip}), viz:
{\small
\beq
\frac{d_L(z)}{d_A(z)(1 + z)^{2}} = (1+z)^{\beta-1} \exp\left(\gamma \int_0^z \frac{dz'}{E(z')(1+z')^{\alpha}}\right)
\label{3param}
\eeq}
where $E(z) \equiv H(z)/H_0$ is the dimensionless Hubble expansion normalised to unity today. Distance-duality corresponds to $(\beta, \gamma) = (1,0)$ (in which case $\alpha$ is arbitrary). 

\subsection{Lensing and finite beams}\label{lensing}

Distance-duality holds exactly only for infinitesimal light bundles in which case gravitational lensing has no effect on the duality. In practice however, observations are made with different finite-sized bundles. Estimates based on observations on large angular scales, (such as the 2df 10QZ survey \cite{outram} or the proposed KAOS survey\footnote{See http://www.noao.edu/kaos/}) will be very weakly affected by gravitational lensing, while SNIa observations may be strongly affected by lensing (by an amount up to $0.3$ mag \cite{1997ff,19972} or more), depending on the fraction of compact objects in the universe. Using such different techniques to estimate $d_A$ and $d_L$ implies that lensing will violate distance-duality by an amount that depends on the fraction of compact objects \cite{ar}. This opens the interesting possibility that future data will be able to test the fraction of compact objects by searching for such lensing-induced violations of distance-duality.

One way to get around this lensing-induced violation is to analyse objects that can give both $d_L$ and $d_A$. An interesting possibility in this category are type 2 SN where $d_A$ can be estimated from observations of the photosphere. Unfortunately $d_A$ data of this sort is currently limited to very low redshift 
\cite{type2}.

\section{Constraints from Current Data}
 
Here we use the standard FLRW equations to calculate the theoretical distance $d_A(z)$ as a function of the cosmic parameters $(\Omega_M,\Omega_{\Lambda})$ (over which we then marginalise, as they are determined by the angular diameter distance data) and use (\ref{3param}) to infer $d_L(z)$ given $(\alpha,\beta,\gamma)$. 

We use a standard Markov-Chain Monte Carlo method with
one chain of $10^6$ points per model to sample the likelihood
and derive the marginalised limits. It is probably useful to
point out here that the 3-parameter eq.~\ref{3param}
contains two ``artificial'' degeneracies clearly visible
in the likelihood contours of Fig (\ref{fig2}) and (\ref{fig3}): 
firstly if $\gamma=0$ then $\alpha$
is completely unconstrained (and $\beta \approx 1$) and secondly
there is a value of $\alpha$ around $0$ for which the
integral is very close to a logarithm, and so the full right
hand side of the equation becomes approximately $(1+z)^{\beta+\gamma-1}$
which leads to the strong degeneracy visible in fig.~\ref{fig3}. More details of our data sets and method are given in appendix A.

\begin{figure}
\begin{center}
\includegraphics[width=80mm]{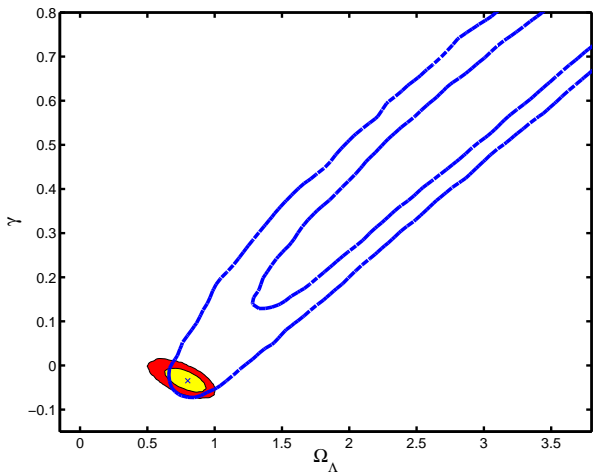} \\
\caption{{\bf Supernovae are brighter relative to $d_A$ data}. $\gamma$-$\Omega_{\Lambda}$ likelihood plot in the case $\alpha = -2, \beta=1$ which corresponds to a photon scattering probability $\propto (1+z)^3$. The best fit corresponds to
$\gamma=-0.036$, i.e. brightening of SNIa relative to the $d_A$ data, as required from figure 1. The very extended, diagonal, contours are the weak $1$ and $2 \sigma$ constraints found using only the SNIa data. This illustrates the power of blending $d_L$ and $d_A$ data as a consistency check of existing data and as a test of new physics.}
\label{scat}
\end{center}
\end{figure}
For this reason it is preferable to study the absolute goodness
of the fit in the full three-dimensional parameter space of
$(\alpha,\beta,\gamma)$. Hence instead of marginalising we show plots 
found by maximising (equivalent to marginalisation in the case of 
a Gaussian likelihood). In this way, it is
easier to see where the well-fitting models are located. When
quoting the limits on $\gamma$ we do
of course marginalise. 

In Figure (1) we show the binned $d_L(z)$ and $d_A(z)$ data as a function of redshift converted to magnitude (relative to the flat concordance model) assuming distance-duality holds, in which case both data sets {\em should lie on the same curve}. The shaded regions shows the effect of the best-fit $\gamma = -0.036$ ($\alpha=-2,\beta=1$) on the underlying $d_A(z)$ showing how it is possible to simultaneously fit the $d_L$ and $d_A$ data with a single model. Also shown are the very different best-fits to the $d_L$ and $d_A$ data taken separately. While the $d_A$ data favour a flat universe, the SNIa data favour a very closed model (ruled out from the CMB) due to the unexpected brightening at $z>0.5$.

Figures (3) \& (4) show the results of our Monte-Carlo Markov Chain likelihood analysis.  It is clear that the distance-duality prediction ($\beta =1, \gamma = 0$) is not favoured by current data with the best fit occurring in the degeneracy region at the edge of the figure, $(\alpha, \beta, \gamma) = (0.1, 4.0, -2.7)$ with $\chi^2_{min} = 217$. In comparison, the best-fit FLRW model, $(\beta, \gamma) = (1, 0)$, has $\chi^2_{min} = 223$.

Figure (2) shows the joint $\gamma-\Omega_{\Lambda}$ likelihood that follows when one imposes $\beta=1$ and $\alpha = -2$ by assuming scattering from objects whose number density scales as $(1+z)^3$ (such as Compton scattering by free-electrons) we find that the best fit for the absorption coefficient is 
$\gamma = -0.036$ with $\chi^2_{min} = 219$ and $-0.07 < \gamma < 0$ at 95\% confidence. Surprisingly, the best-fit corresponds not to absorption but brightening, as is clear from figure (1) since the $d_A(z)$ data lies above the $d_L$ points. The extra parameters are justified in all cases from the Akaike information criterion while the Bayesian information criterion favours introducing $\gamma$ at the expense of curvature, while it marginally disfavours introducing all of $(\alpha, \beta, \gamma)$ depending on the binning of the data (see e.g. \cite{andrew}).
\begin{figure}
\begin{center}
\includegraphics[width=80mm]{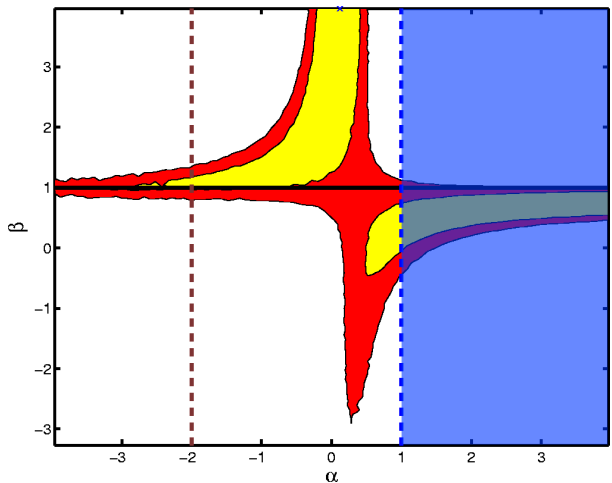} \\
\caption{{\bf Goodness-of-fit contours (1 and 2 $\sigma$) for the parameters $\alpha$ and
$\beta$ of equation (\ref{3param})}. Distance-duality implies $\beta=1$. If photons decay at a constant rate, then $\alpha=1$. Higher
values of $\alpha$ correspond to models where the photon scattering cross section increases as the
universe expands. If photons are affected by ``scatterers'' with a constant comoving density, then $\alpha=-2$. $\alpha > 1$ corresponds to a rather unphysical region of parameter space where the 
probability of photons scattering {\em increases} with the expansion of the universe.
\label{fig2}}\end{center}
\end{figure}
The magnitude of the effect corresponds to an increase of about $5\%$ in the number of photons per Hubble time, a very large violation of photon conservation. We can put this into perspective by comparing it with the expected loss of photons due to Compton scattering by the free electrons in the ionised inter-galactic medium. At $z < 3$ helium is expected to be doubly ionised ($f_Y=0.5$), leading to a free-electron density $n_e = \Omega_b \rho_{crit}(1- Y f_Y)/m_N$ where $Y=0.24$ is the primordial helium abundance. We therefore find a scattering amplitude of $\gamma_{Compton} = \sigma_T n_e/(2 H_0) \sim 10^{-3}$, a factor of about $50$ less than the best-fit (and of opposite sign). 

A plausible explanation is magnification bias through gravitational lensing. If distant SNIa are preferentially  detected if they are brightened then this would cause an apparent violation of the reciprocity relation as discussed in section (\ref{lensing}). It has recently been demonstrated that lensing does significantly affect current high-z SNIa samples \cite{WS}. Brighter high-z SNIa are preferentially found behind overdense regions of galaxies and can differ from demagnified SNIa by $0.3-0.4 mag$. The induced bias may be sufficient to provide the $\sim 0.1-0.2 mag$ brightening required to remove the violation of distance-duality we have documented above. Alternatively, since smaller compact radio sources are typically brighter \cite{jackson}, in an incomplete magnitude limited survey, high-redshift sources will be systematically smaller and yield a larger value of $d_A(z)$.

\subsection{Ruling out replenishing dust}

Riess {\em et al} \cite{riess} found that the best-fit model to all currently available SNIa was not an accelerating $\Lambda$CDM model but rather a replenishing grey-dust model \cite{goobar} with $\Lambda = 0$ which causes redshift-dependent dimming of the SNIa, with $\alpha$ changing from $-2$ to $1$ at $z=0.5$. If this was the correct explanation then we should expect a marked violation of distance duality with the $d_A$ data lying below the $d_L$ data since it would correspond to a non-accelerating universe. Our results show that this is not the case (indeed we have the opposite problem!)

A detailed analysis of this model based on \cite{goobar,ccm} gives a best-fit to {\em all} the data of 
$\Omega_{\Lambda} = 0.77 \pm 0.13$ showing that the combined data, in contrast to the 
SNIa data alone, rule out the replenishing dust model at over 4-$\sigma$.
\begin{figure}\begin{center}
\includegraphics[width=80mm]{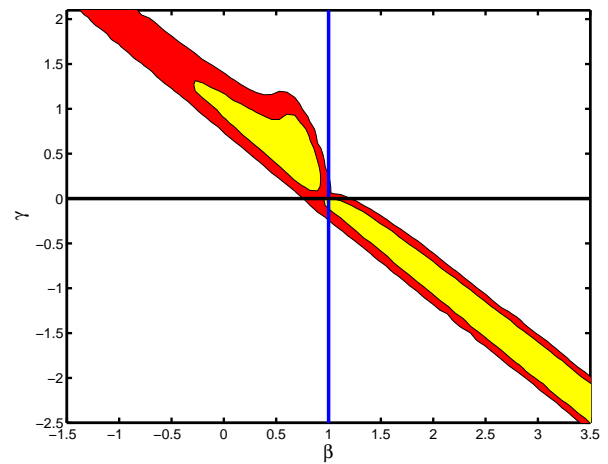} \\
\caption{{\bf Goodness-of-fit contours (1 and 2 $\sigma$) for the parameters $\alpha$ and
$\beta$ of equation (\ref{3param})}. Distance-duality implies $\beta=1$ and $\gamma=0$, which corresponds to photon conservation. This point is acceptable for the three-parameter case, but becomes unfavoured when we limit ourselves to the sub-space $(\beta=1,\alpha=-2)$ of constant comoving-density scatterers.
}\label{fig3}\end{center}
\end{figure}
\subsection{Is dark energy evolving?}

While we have discussed interesting physics which violates distance-duality, dark energy dynamics is not among them. Hence evidence from SNIa for significant evolution in the dark energy equation of state $w(z)$ at low redshift \cite{meta,cond,sahni,riess} now appear less significant since the signal is not seen in the $d_A(z)$ data. As our measurements of distance-duality improve we will be able to obtain better constraints on dark energy evolution. 

\section{Conclusions}

In this paper we have emphasised distance-duality as a test of fundamental and exotic physics related to the metric nature of gravity and photon conservation on cosmic scales.  Although stringent constraints will arise in the next few years the test is already proving powerful. In particular we are able to essentially rule-out non-accelerating models of the universe which explain the supernova dimming by grey-dust scattering, extinction or evolution. Interestingly, current data suggests a small (2-$\sigma$) discrepancy that may be due to lensing-induced magnification bias of the high-z SNIa.

One can ask if it will be possible to distinguish violation of photon conservation from deviations of gravity from a metric theory assuming systematics errors are eliminated. One interesting way to do this would be to use binary black holes as standard gravitational wave candles \cite{gw} to give an independent estimate of $d_L(z)$. Comparing this against the $d_L(z)$ found from SNIa using e.g. the JDEM/SNAP satellite and against distance-duality should allow us to distinguish between the two possibilities.  Further, large galaxy surveys such as the proposed KAOS experiment, will provide accurate estimates of $d_A(z)$ out to $z=3$ \cite{se},  allowing us to test deviations from distance-duality at the level of a few percent, implying that this diagnostic will mature into a unique and powerful test of fundamental physics on cosmological scales. 

We thank Tom Andrews, Sarah Bridle, Robert Caldwell, Pier-Stefano Corasaniti, George Ellis, Ariel Goobar, Roy Maartens, Peter Nugent, Misao Sasaki, Takahiro Tanaka and Licia Verde for useful discussions, John Tonry for discussions and his SNIa likelihood code and Adam Riess for discussions and the missing supernova. BB is supported by the Royal Society \& JSPS and thanks UCT for hospitality. MK is supported by PPARC. 

\appendix
\section{Data set details}

The main supernova data set $(d_A)$ is the 
``gold'' subset of  
Riess {\em et al} \cite{riess}. We checked that this gives essentially the same 
results as the earlier data in
Tonry {\em et al.} \cite{tonry}, Barris {\em et al.} \cite{barris},
and Knop {\em et al.} \cite{knop}\footnote{
We used the extinction corrected data ($m_B^{\rm eff}$ in
table 3 of \cite{knop}), and as we use only their new supernovae we
cannot easily apply the stretch correction. Hence this
data set does not improve the SNIa constraints much.} (TBK).
Although we used our own code to evaluate the resulting likelihood, 
it follows closely the one of John Tonry 
and gives the same results.

For the $d_A$ estimates we used the data sets of
Daly \& Djorgovski \cite{daly1} (DD), Gurvits \cite{crg} (G) and
Jackson \cite{jackson} (J). DD provide their data directly as
dimensionless $y(z)$ and we use it in this form. G gives the data as
angular sizes $\theta(z)$ with $d_A = l/\theta$ and we need to
marginalise over the unknown ``standard ruler'' $l$. This is
analogous to the case of supernovae. For this reason 
the radio galaxy data also does not depend on the Hubble constant.
J also provides angular sizes, but pre-binned and uses 
error bars determined so that the
resulting $\chi^2$ value per degree of freedom is unity. We then
marginalise over an independent angular size $l'$ in this case as
well. We checked that we obtain the same confidence regions as
\cite{jackson} when using the J data set alone.

How stable are our results to changes of the underlying data sets?
Taking the absorption model as a test case, 
leaving out any single $d_A$ data set does not 
change the constraints on $\gamma$ appreciably, although if we
drop DD then $\gamma=0$ becomes acceptable at $2 \sigma$.
If on the other hand we use the combined supernova data sets
of TBK, we find stronger evidence for a violation of
distance-duality, with $\gamma < -0.01$ at $2 \sigma$.
The conclusion that there is something systematically 
different between the SNIa and the RG data sets is therefore 
rather stable.

As a further test of the radio galaxy data,
we have included the gas mass fraction data
of Allen {\em et al.}, and marginalised over all nuisance parameters
(in this case the bias, the Hubble constant and the baryon
density). The X-ray data is
consistent with the RG data, and its addition does not change the
constraints on $\gamma$. But as eq. (15) in \cite{allen}
is only given for flat universes, we quote our results without this data set.
Strong constraints may also come in the future from SZ data \cite{reese}. 


\end{document}